\documentclass[reprint,aps,prb,amsmath,amssymb,superscriptaddress]{revtex4-2}
\usepackage{graphicx}

\begin{document}

\title{Spin and charge excitations in the correlated multiband metal Ca$_3$Ru$_2$O$_7$}

\author{J.~Bertinshaw}
\email{j.bertinshaw@fkf.mpg.de}
\affiliation{Max Planck Institute for Solid State Research, Heisenbergstr.~1, D-70569 Stuttgart, Germany}
\affiliation{Deutsches Elektronen-Synchrotron DESY, Notkestr. 85, D-22607 Hamburg, Germany}
\author{M.~Krautloher}
\author{H.~Suzuki}
\author{H.~Takahashi} 
\affiliation{Max Planck Institute for Solid State Research, Heisenbergstr.~1, D-70569 Stuttgart, Germany}
\author{A. Ivanov}
\affiliation{Institut Laue-Langevin, 71 avenue des Martyrs, CS~20156, 38042~Grenoble~cedex~9, France}
\author{B.~J.~Kim}
\affiliation{Max Planck Institute for Solid State Research, Heisenbergstr.~1, D-70569 Stuttgart, Germany}
\affiliation{Department of Physics, Pohang University of Science and Technology, Pohang 790-784, South Korea}
\affiliation{Center for Artificial Low Dimensional Electronic Systems, Institute for Basic Science (IBS), 77 Cheongam-Ro, Pohang 790-784, South Korea}
\author{H.~Gretarsson}
\email{hlynur.gretarsson@desy.de}
\affiliation{Deutsches Elektronen-Synchrotron DESY, Notkestr. 85, D-22607 Hamburg, Germany}
\author{B.~Keimer}
\affiliation{Max Planck Institute for Solid State Research, Heisenbergstr.~1, D-70569 Stuttgart, Germany}

\date{\today}

\begin{abstract}
We use Ru $L_3$-edge resonant inelastic x-ray scattering (RIXS) to study the full range of excitations in Ca$_3$Ru$_2$O$_7$ from meV-scale magnetic dynamics through to the eV-scale interband transitions. This bilayer $4d$-electron correlated metal expresses a rich phase diagram, displaying long range magnetic order below 56~K followed by a concomitant structural, magnetic and electronic transition at 48~K. In the low temperature phase we observe a magnetic excitation with a bandwidth of $\sim$30~meV and a gap of $\sim$8~meV at the zone center, in excellent agreement with inelastic neutron scattering data. The dispersion can be modeled using a Heisenberg Hamiltonian for a bilayer \mbox{$\mathrm{S}=1$} system with single ion anisotropy terms. At a higher energy loss, $dd$-type excitations show heavy damping in the presence of itinerant electrons, giving rise to a fluorescence-like signal appearing between the $t_{2g}$ and $e_g$ bands. At the same time, we observe a resonance originating from localized $t_{2g}$ excitations, in analogy to the structurally related Mott-insulator Ca$_2$RuO$_4$. But whereas Ca$_2$RuO$_4$ shows sharp separate spin-orbit excitations and Hund's-rule driven spin-state transitions, here we identify only a single broad asymmetric feature. These results indicate that local intra-ionic interactions underlie the correlated physics in Ca$_3$Ru$_2$O$_7$, even as the excitations become strongly mixed in the presence of itinerant electrons.
\end{abstract}

\maketitle
\section{Introduction}
\begin{figure*}
    \includegraphics{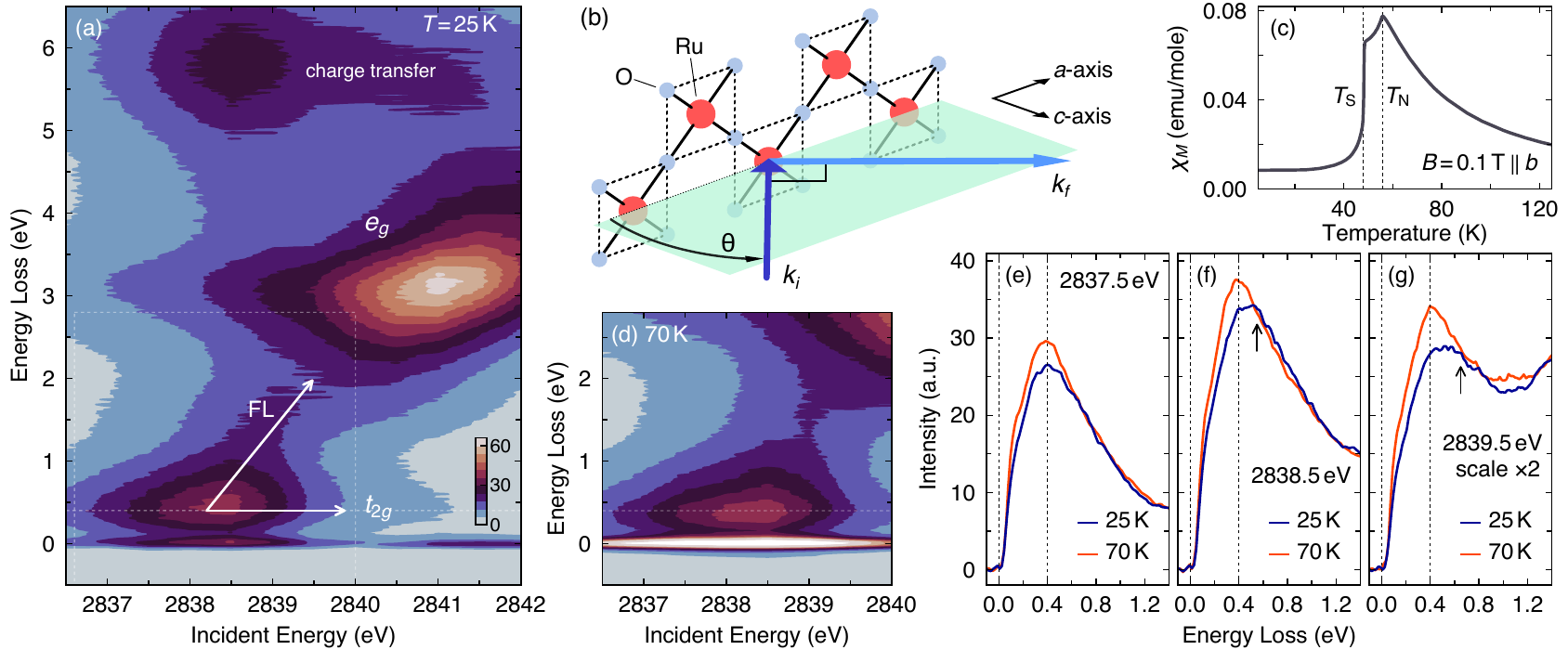}
    \caption{\label{fig:rixs}
    a) RIXS energy map taken at \mbox{$T=25$~K} plots the incident-energy dependence of the energy loss spectra across the Ru $L_3$ edge.
    The quasielastic magnon signal and $t_{2g}$ feature at $\sim$0.4~eV (horizontal dashed line) resonate around \mbox{$E_i=2838.5$~eV}, while $e_g$ excitations around 3~eV strongly resonate at 2841~eV.
    At higher energies ($\sim$6~eV) charge transfer excitations are present.
    b) The sample was oriented such that the Ru-O-Ru bonds were aligned $45^\circ$ to the scattering plane. Outgoing photons $k_f$ were detected at a fixed angle of $90^\circ$ with respect to the incoming beam $k_i$. The momentum-transfer was selected by varying the angle $\theta$ between $k_i$ and the RuO$_2$ planes.
    c) Bulk magnetic response of Ca$_3$Ru$_2$O$_7$, with \mbox{$B=0.1$~T} applied along the $b$-axis.
    d) RIXS energy map conducted above $T_\mathrm{N}$ (region indicated by a dashed rectangle in (a)) shows a dramatic enhancement of the quasielastic magnetic excitation and an apparent shift of the $t_{2g}$ spectral weight to smaller energy loss.
    e-g) Comparison of individual RIXS spectra from (a) and (d). The quasielastic contribution has been removed. Dashed vertical line at 0.4~eV lies at the $t_{2g}$ maximum at $T=70$ K. The arrows identify a second mode with an independent incidence-energy response.
    }
    \end{figure*}

Ca$_3$Ru$_2$O$_7$ is a canonical multiband correlated metal, displaying complex behavior and a rich phase diagram despite the lack of localized $d$-orbital valence electrons generally present in strongly correlated oxides~\cite{ Ohmi04, Baum06, Bao08}. The interactions that drive correlated phenomena away from the Mott-state in $4d$-electron Ca$_3$Ru$_2$O$_7$ and other multiband metals like superconducting Sr$_2$RuO$_4$ and the iron-pnictides are an active and ongoing concern~\cite{Geor13}. Vital to this understanding is an investigation of the electronic excitations, which encode information regarding the energy scales of intra- and inter-ionic interaction parameters. Here, resonant inelastic x-ray scattering (RIXS) at the dipole-active Ru $L_3$-edge (\mbox{$2p \rightarrow 4d$}) has emerged as a key tool, providing a direct momentum and energy-dependent probe of electronic excitations~\cite{Gret19, Suzu20}. Moreover, Ru $L_3$-edge RIXS also covers magnetic transitions, giving detailed insight into spin wave dynamics~\cite{Suzu19}. In a recent RIXS study of the single layered antiferromagnetic Mott-insulator Ca$_2$RuO$_4$, clear spin-orbital and Hund's-rule driven intra-ionic excitations were identified, from which the spin-orbit coupling (SOC) $\xi$, the Hund's rule energy $J_{\mathrm{H}}$, and the tetragonal crystal field term $\Delta$ were extracted~\cite{Gret19}.

The bilayer system Ca$_3$Ru$_2$O$_7$ retains a highly anisotropic electric resistivity down to lowest temperatures~\cite{Yosh04}, where the optical conductivity reveals a small pseudogap of 25~meV~\cite{Lee07}. The magnetic properties of the system are dominated by intra-bilayer ferromagnetic exchange~\cite{Yosh05}. Long range magnetic order forms below \mbox{$T_{\mathrm{N}}=56$~K} where ferromagnetic bilayers stack antiparallel along the $c$-axis (AFM-$a$)~\cite{Ke11a}. At \mbox{$T_{\mathrm{S}}=48$~K} a structural transition that distorts the RuO$_6$ octahedra coincides with an upturn in the out-of-plane resistivity~\cite{Ohmi04, Yosh05}, and a spin rotation from the $a$-axis to $b$-axis (AFM-$b$) that appears to be mediated by an incommensurate spin state~\cite{Yosh04, Bohn08, Dash20}. The system also shows a complex magnetic field dependence~\cite{Bao08, Soko19}, and a remarkable response to doping, where dilute substitution of non-magnetic $3d^0$-electron Ti$^{4+}$ on the Ru$^{4+}$ site quickly pushes the system into a Mott state with antiferromagnetic dynamics characteristic of Ca$_2$RuO$_4$~\cite{Tsud13, Krau20}. Taken together, these results signify the presence of a delicate balance of competing interactions, which can be captured in spectroscopic studies at energies beyond the reported small pseudogap.

In this report we present a systematic Ru $L_3$-edge RIXS study of Ca$_3$Ru$_2$O$_7$. Below $T_{\mathrm{S}}$ we capture the dispersion of the in-plane magnon across the entire Brillouin zone, giving a spin-wave gap of $\sim$8~meV, in excellent agreement with inelastic neutron scattering (INS), and a bandwidth of $\sim$30~meV. At higher energies we identify dispersionless $dd$-excitations broadened by itinerant electrons. By studying the RIXS response as a function of incident-energy and temperature we conclude that localized and fluorescence-like excitations are found simultaneously within the $t_{2g}$ multiplet. A comparison with the structurally related Ca$_2$RuO$_4$ suggests that the localized modes represent intra-ionic spin-orbit excitations and Hund's multiplets, which become heavily mixed by the electronic continuum present in Ca$_3$Ru$_2$O$_7$.

\section{Experimental Details}
Single crystals of Ca$_3$Ru$_2$O$_7$ and Ca$_2$RuO$_4$ were grown using a floating zone method described previously~\cite{Naka01}. High quality detwinned samples were identified and aligned using polarized light microscopy~\footnote{See Supplemental Material for polarized light microscopy and INS spectra} and magnetometry. RIXS measurements were carried out at beamline P01 at the PETRA-III synchrotron at DESY, using the IRIXS spectrometer~\cite{Gret20}. A cryogenically cooled Si(111) two-bounce monochromator, secondary Si(111) four-bounce monochromator (asymmetrically cut) and focusing KB-mirror optics were used in combination with a spherically diced SiO$_2$ (10$\bar{2}$) analyzer to obtain an overall energy resolution \mbox{$\Delta \mathrm{E} \sim 75$~meV full-width at half-maximum}. To determine the energy of the elastic line, we measured scattering from a droplet of GE varnish applied to the corner of the sample. The RIXS studies were carried out in a \mbox{$(\mathrm{H}00)\times(00\mathrm{L})$} scattering geometry (orthorhombic unit cell) as depicted in Fig.~\ref{fig:rixs}b).

\section{Results}
\subsection{Incident energy dependence}
In Fig.~\ref{fig:rixs}a) a RIXS incident energy ($E_i$) map of Ca$_3$Ru$_2$O$_7$ is shown, collected around the Ru $L_3$-edge \mbox{($\sim$2840~eV)}. The spectra are not normalized, and only nominal elastic scattering is present, indicative of the high quality of the crystal. The sample was cooled through the magnetically ordered phase to \mbox{$T=25\,\mathrm{K}<T_{\mathrm{N}}$} and orientated with an incident angle \mbox{$\theta=45^\circ$} such that spectra were collected at the Brillouin zone center $\Gamma$ (\mbox{$Q_{\mathrm{HKL}}=(00\mathrm{L})$}). Following the process described in Ref.~\onlinecite{Gret19} for Ca$_2$RuO$_4$, we consider the spectra as a series of components: a low-energy quasielastic peak followed by electronic $dd$-excitations originating from $t_{2g}$ states below 1.5~eV and \mbox{$t_{2g} \rightarrow e_{g}$} above 2~eV. In addition, spectral weight forms at intermediate energies that connects the $t_{2g}$ and $e_{g}$ features. Since the energy loss of these spectral weights changes with the incident energy we assign it to a fluorescence-like response. In other words, it originates from delocalized electronic excitations due to the virtually ungapped electron-hole continuum in Ca$_3$Ru$_2$O$_7$. This suggests coexisting excitations with Raman-like (i.e., excitation energy independent of $E_i$) and fluorescent behavior (i.e., excitation energy following $E_i$), which has also been reported in recent soft x-ray RIXS experiments on nickel oxides~\cite{Biso16}, copper oxides~\cite{Mino17} and iron arsenides~\cite{Gilm20}.

As charge degrees of freedom in Ca$_3$Ru$_2$O$_7$ remain below the energy of the lowest $dd$-excitation, the quasielastic scattering can originate from either spin or charge transitions. In order to identify the origin of the quasielastic scattering, an incident energy map was therefore repeated in the paramagnetic phase at \mbox{$T=70\,\mathrm{K}>T_{\mathrm{N}}$}, as shown in Fig.~\ref{fig:rixs}d). The quasielastic resonance undergoes a dramatic enhancement in the paramagnetic state, even as the intensity of the higher-energy signal remains comparatively unaffected, confirming the magnetic nature of the peak. We note that the intensity remains peaked at $\Gamma$ above $T_{\mathrm{N}}$, indicating that the fluctuations remain ferromagnetic in the paramagnetic regime.

Moving to the $t_{2g}$ excitation around 0.4~eV, marked with horizontal lines in Fig.~\ref{fig:rixs}a) and d), a distinct change in the response is seen as the spectral weight shifts towards lower energies at \mbox{$T=70$~K}. This change in the $t_{2g}$ excitation is made clear in Fig.~\ref{fig:rixs}e-g), where the temperature differences for three incidence energies are compared (the quasielastic magnetic contribution has been subtracted for clarity). Here a complex incident energy dependence emerges. In the paramagnetic phase the central weight of the $t_{2g}$-excitation remains fixed around 0.4~eV (vertical dashed lines in Fig.~\ref{fig:rixs}e-g)) irrespective of $E_i$. In the magnetically ordered phase, the overall spectral weight decreases and an energy dependence emerges -- spectral weight starts to shift towards higher energies as $E_i$ increases (see arrow at 0.6~eV in Fig.~\ref{fig:rixs}f) and g).

\subsection{Comparison with Ca$_2$RuO$_4$}
\begin{figure}
    \includegraphics{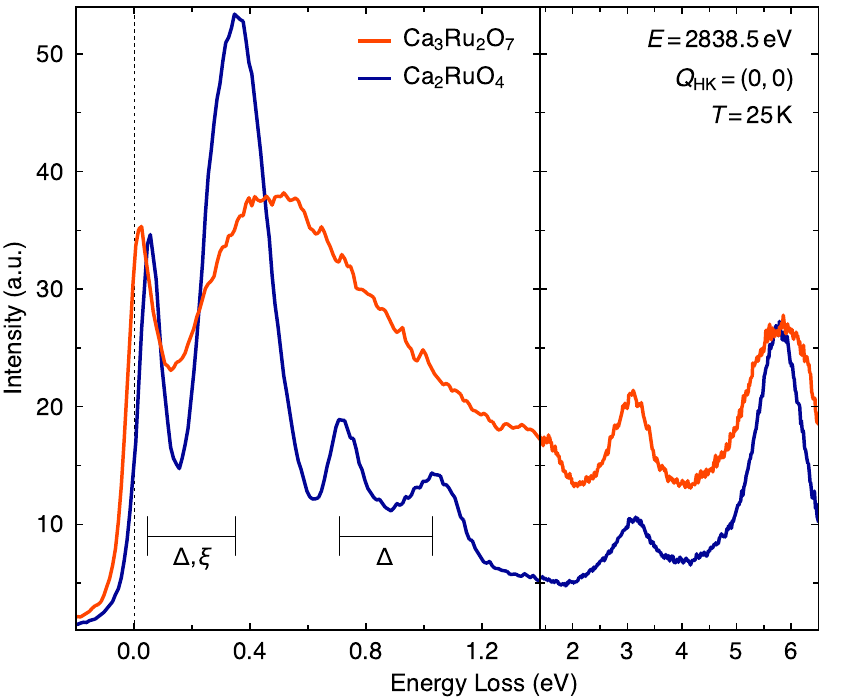}
    \caption{\label{fig:comp}
        Ca$_3$Ru$_2$O$_7$ RIXS spectrum, measured at \mbox{$T=25$~K} and \mbox{$E_i=2838.5$~eV} at \mbox{$Q_{\mathrm{HKL}}=(00L)$} \mbox{($\theta=45^\circ$)}, is compared with Ca$_2$RuO$_4$. The Ca$_2$RuO$_4$ data is scaled by a factor of 0.7.
        Modes associated with spin-wave, $t_{2g}$ and $e_g$ excitations as well as charge-transfer are present in both samples.
        The low-energy splitting of the excitations in Ca$_2$RuO$_4$ are associated with spin-orbit coupling and tetragonal crystal field terms.}
        \end{figure}

RIXS spectra of Ca$_3$Ru$_2$O$_7$ and the structurally related Mott-insulator Ca$_2$RuO$_4$ are compared in Fig.~\ref{fig:comp}. Both systems were measured under the same conditions at the $t_{2g}$ resonance \mbox{$E_i=2838.5$~eV} (the Ca$_2$RuO$_4$ spectrum is scaled by a factor of 0.7 as a visual aid). At higher energies both Ca$_3$Ru$_2$O$_7$ and Ca$_2$RuO$_4$ show a similar signal, with charge transfer type excitations at 6~eV and {$t_{2g}\rightarrow e_g$} excitations at 3~eV. It is perhaps not surprising that there are no significant changes within this energy regime, given that the overall RuO$_6$ symmetry and Ru$^{4+}$ and O$^{2-}$ valencies are identical in the two systems. Below 0.1~eV the magnetic excitation in Ca$_2$RuO$_4$ is located higher in energy than Ca$_3$Ru$_2$O$_7$; a comparison of the spin waves is discussed in the following section.

The largest difference arises within the $t_{2g}$ regime below $\sim$1.5~eV, where the broad asymmetric feature in Ca$_3$Ru$_2$O$_7$, which we note has a striking resemblance to Sr$_2$RuO$_4$~\cite{Fatu15}, contrasts strongly with the series of sharp excitations observed in Ca$_2$RuO$_4$. The excitations in Ca$_3$Ru$_2$O$_7$ are broadened presumably through coupling with the electronic continuum present at all energies due to the lack of a charge gap. At the same time, the shift in spectral weight to higher energy loss implies a reconfiguration of the $t_{2g}$ multiplet structure. The energies of the $t_{2g}$ modes in Ca$_2$RuO$_4$---a \mbox{$\mathrm{J}=0\rightarrow 2$} spin-orbit excitation at 0.32~eV followed by Hund's-rule driven \mbox{$\mathrm{S}=1\rightarrow 0$} spin-state transitions around 0.75 and 1.0~eV---are controlled primarily by the $\Delta$, $\xi$, and $J_{\mathrm{H}}$ parameters~\cite{Gret19}. In particular, the large splitting between the $\mathrm{S}=1\rightarrow 0$ excitations directly reflects the magnitude of $\Delta$ (see labels in Fig.~\ref{fig:comp}). In Ca$_3$Ru$_2$O$_7$ the RuO$_6$ octahedra are less compressed~\cite{Brad98, Yosh05}, which will have the effect of reducing $\Delta$. In the ionic picture this leads to a smaller splitting of the \mbox{$\mathrm{S}=1\rightarrow 0$} excitations and lowering of the  $\mathrm{J}=0\rightarrow 2$ excitation. As such, the observed dichotomy in the energy-dependency response in Fig.~\ref{fig:rixs}d-f) indicates that underlying the asymmetric $t_{2g}$ profile are excitations associated with the ionic model, which are heavily mixed with other charge degrees of freedom but remain distinct from the metallic continuum response of the system.

\subsection{Magnetic dispersion}
\begin{figure}
    \includegraphics{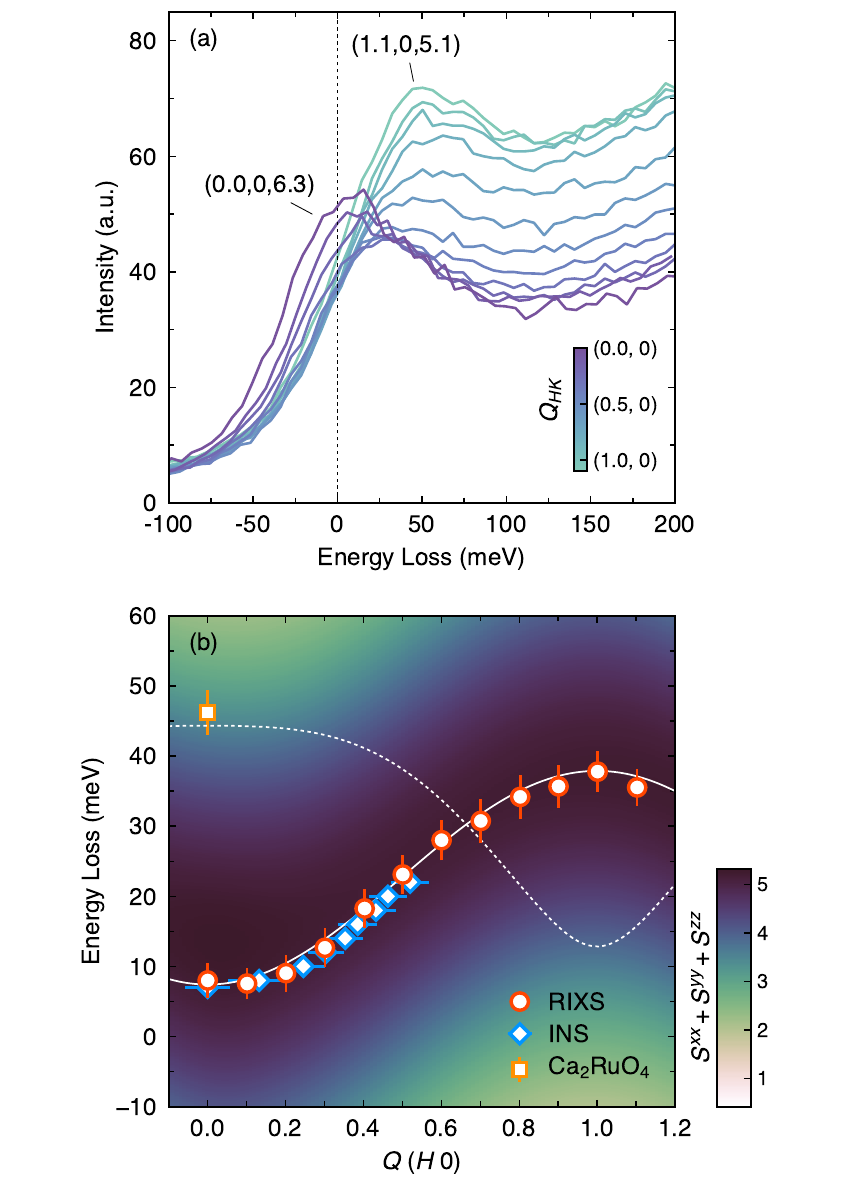}
    \caption{\label{fig:mag}
        a) The momentum dependence of the RIXS signal below \mbox{$E=200$~meV} at \mbox{$T=25$K} from zone center to zone boundary shows the dispersion of the magnon.
        b) The fitted magnon dispersion captured with RIXS corresponds well with inelastic neutron spectroscopy. Solid white lines plot the acoustic and optical modes of the bilayer spin wave model as discussed in the text, while the colormap represents the intensity calculated from the spin-spin correlation function that takes into account instrumental momentum and energy resolutions. The Ca$_2$RuO$_4$ spin wave position determined from Fig.~\ref{fig:comp}, and model spin wave based on Ref.~\onlinecite{Jain17} (dashed white line) are shown for comparison.
        }
        \end{figure}

The lowest energy regime associated with magnetic excitations was studied in a detailed $Q$-dependence to follow the spin wave from the zone center \mbox{$Q_{\mathrm{HKL}}=(0,0,6.3)$} to slightly beyond the zone boundary $(1.1,0,5.1)$. RIXS spectra in the region close to the elastic line are plotted in Fig.~\ref{fig:mag}a), which shows the clear dispersion of the spin wave from the zone center to boundary. The intensity is enhanced close to \mbox{$Q_{H}=0.0$}, consistent with ferromagnetic coupling within the bilayers. The fitted position of the magnon across the Brillouin zone is plotted in Fig.~\ref{fig:mag}b), which was extracted using a fitting procedure described in the following section. It is overlaid with the low-energy regime of the spin wave determined using inelastic neutron scattering, which was collected using spectrometer IN8, ILL, Grenoble, in the FlatCone configuration~\cite{Kemp06, IN8data, Note1}. We note that while the full dispersion could be extracted from the RIXS spectra, the neutron scattering signal became too weak at higher energies to study the spin wave (See also Ref.~\onlinecite{Ke11a}). Despite this, the striking similarity of the system's response to RIXS and INS clearly illustrates that the two approaches are complementary probes of the same underlying magnetic dynamics.

A minimal Heisenberg Hamiltonian for a bilayer $S=1$ system takes into account in-plane superexchange coupling ($J$) between Ru moments $\boldsymbol{S}$ as well as the intra-bilayer exchange interaction ($J_C$) between directly adjacent moments stacked along the c-axis. Following prior work on Ca$_2$RuO$_4$~\cite{Jain17}, we also introduce tetragonal ($E$) and orthorhombic ($\epsilon$) single-ion anisotropy (SIA) terms to account for the spin wave gap:
\begin{align*}
    \label{eq:hamiltonian}
    H = J \sum_{\langle i,j \rangle}
    \boldsymbol{S}_i & \cdot \boldsymbol{S}_j 
    + J_{c} \sum_{\langle i,j \rangle_c}
    \boldsymbol{S}_i \cdot \boldsymbol{S}_j
  \\+ &E \sum_i S^{z}_{i}{}^2
  + \epsilon \sum_i S^{x}_{i}{}^2.
\end{align*}

Coupling between bilayers, although important to the bulk antiferromagnetic response, is exceedingly weak and neglected here. Within the bilayer structure we would expect to identify acoustic and optical spin wave modes associated with the respective in-plane and intra-bilayer couplings~\cite{Note1}. The bilayer structure factor gives rise to an intensity modulation of the two modes, resulting in a maximum intensity of the in-plane (out-of-plane) mode at \mbox{$Q_{\mathrm{L}}=5.0\,(7.5)$} and minimum at \mbox{$Q_{\mathrm{L}}=7.5\,(5.0)$}. As such, while the majority of the intensity is associated with the in-plane mode, the energy resolution limit makes it impossible to separate the two modes. We do however use this mixing to constrain the maximum energy of $J_c$. The primary spin wave for \mbox{$J=-3.75$~meV}, \mbox{$J_c=-6.5$~meV}, \mbox{$E=5.5$~meV} and \mbox{$\epsilon=2.5$~meV} is plotted in Fig.~\ref{fig:mag}b) as a white solid line, which corresponds well with parameters reported by Ke~\emph{et.~al.}~\cite{Ke11a}. The colormap represents the calculated intensity of the excitation spectrum that takes into account the calculated IRIXS instrumental momentum and energy resolutions.

\subsection{Temperature and momentum dependence}
\begin{figure*}
    \includegraphics{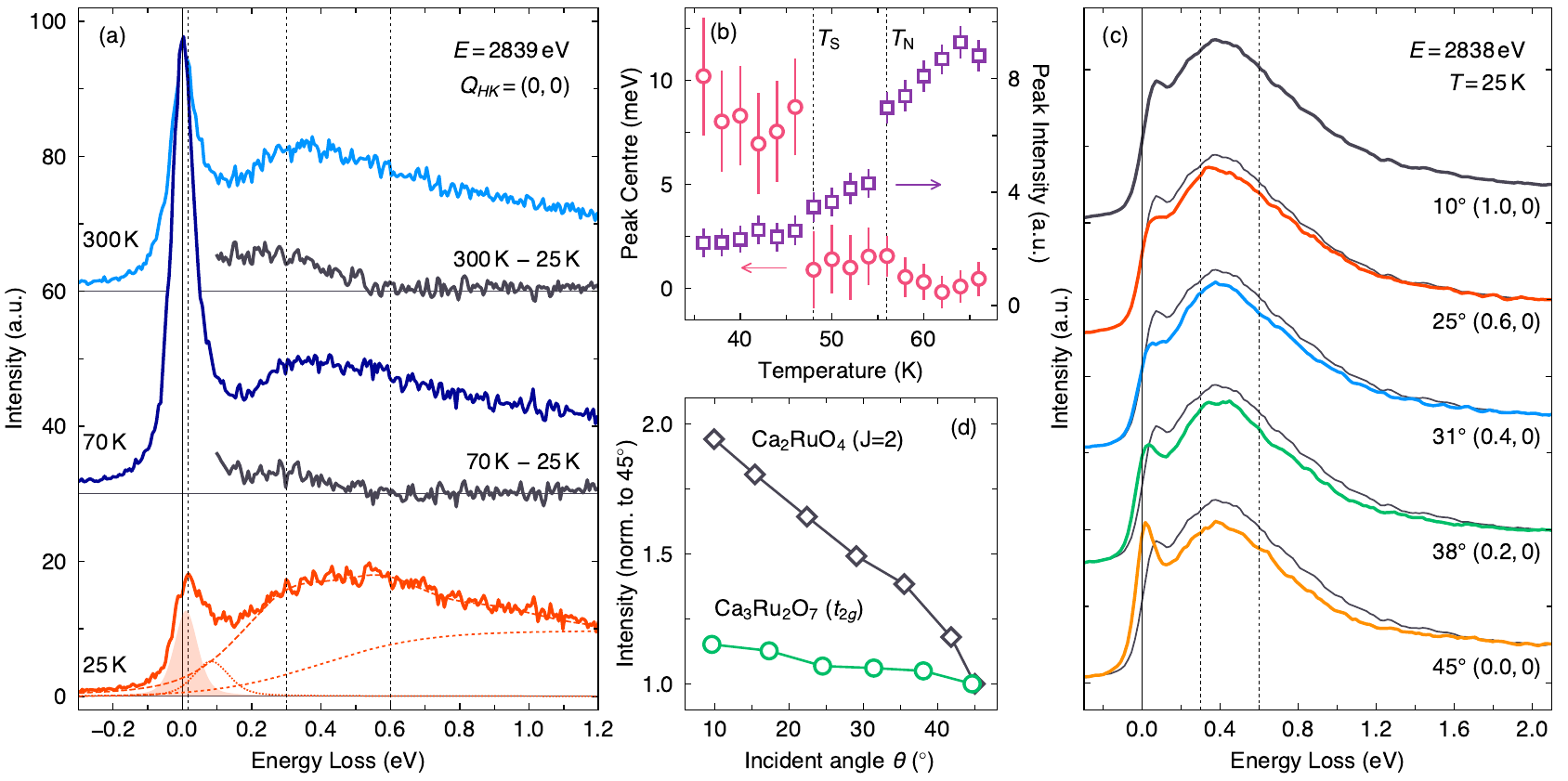}
    \caption{\label{fig:tdep}
        a) Temperature dependence of the magnon at the magnetic zone center, corresponding to the spin wave gap. Above $T_\mathrm{N}$ the gap closes and strongly increases in intensity. At \mbox{$T=300$~K} weak paramagnetic excitations remain.
        b) Fitted magnitude and intensity of the spin wave gap as a function of temperature. Results reveal that the gap closes at $T_\mathrm{S}$, while the bulk of the increase in intensity happens above $T_\mathrm{N}$.
        c) Momentum dependence of the RIXS spectra at \mbox{$T=25$~K} collected by varying $\theta$. Each spectrum was normalized to account for self-absorption and shifted vertically for clarity. Overlaid on each spectrum is the data collected at \mbox{$\theta=10^\circ$}.
        d) Intensity of the $t_{2g}$ features as a function of $\theta$ plotted in comparison to the spin-orbit \mbox{$\textrm{J}=2$} excitation in Ca$_2$RuO$_4$. Only moderate increase in spectral weight is observed in Ca$_3$Ru$_2$O$_7$ at low $\theta$.
        }
        \end{figure*}

We now turn to detailed temperature and momentum-transfer RIXS studies. Figure~\ref{fig:tdep}a) plots the low-energy spectra collected at \mbox{$T=25$~K}, 70~K and 300~K for \mbox{$Q_{\mathrm{HKL}}=(0,0,6.3)$}, which are offset for clarity. Here, spectra were collected at $E_i=2839$~eV in order to enhance the visibility of the separate excitations that arise within the $t_{2g}$ multiplet structure.

Above $T_{\mathrm{S}}$ and $T_{\mathrm{N}}$ at 70~K, the $\sim$8~meV magnon gap, marked by a dashed line in Fig.~\ref{fig:tdep}a), closes as the signal intensity strongly increases. At 300~K ungapped magnetic excitations remain, although the signal is diminished and broadened. To capture the temperature evolution of the magnetic excitations through $T_\mathrm{N}$ and $T_\mathrm{S}$ a series of low-energy spectra were collected at $\Gamma$ from \mbox{$T=35$~K} up to 65~K. The magnon was modeled with a single pseudo-Voigt profile and the electronic continuum with a sigmoid function emerging from the elastic line. Two functions were used to fit the $t_{2g}$ spectral weight. Following the observation in Fig.~\ref{fig:rixs}g) we placed one peak around 600~meV, and the second one around 300~meV (marked by vertical dashed lines in Fig.~\ref{fig:tdep}a)). A small profile around 100~meV was also required to accurately model the excitation spectrum; such a feature was also seen in optical conductivity measurements~\cite{Lee07, Note1}. The resulting fit is shown in Fig.~\ref{fig:tdep}a) for the 25~K dataset, with the magnon profile highlighted and other contributions plotted with dashed lines.

The magnitude of the fitted spin wave gap and the integrated intensity of the magnetic fluctuations are plotted as a function of temperature in Fig.~\ref{fig:tdep}b). The temperature dependency reveals that the base temperature 8(2)~meV magnon gap abruptly forms below the structural transition \mbox{$T_\mathrm{S}=48$~K}, and not the N\'eel temperature \mbox{$T_\mathrm{N}=56$~K}. The step-like behavior of the gap at $T_\mathrm{S}$, which we note coincides with the rotation of the spins from the $a$-axis to the $b$-axis~\cite{Yosh05}, suggests that the magneto-structural transition arises with a marked change in the spin anisotropy of the system. At the same time, the increase in intensity of the magnon through this transition is gradual, as would be expected from a second-order phase transition.

A closer study of the temperature evolution of the electronic $t_{2g}$ multiplet structure was conducted by subtracting the 25~K spectrum from the 70~K and 300~K datasets, which are plotted in Fig.~\ref{fig:tdep}a) as gray lines below the raw spectra.  Here it can be seen that the increase in spectral weight extends up to around 0.5~eV, an energy scale that is much larger than the $<100$~meV pseudogap estimated from optical conductivity~\cite{Lee07} and Raman scattering~\cite{Liu99} measurements. It is therefore clear that components of the electronic structure evolve with temperature significantly away from the Fermi level.

Figure~\ref{fig:tdep}c) plots the momentum dependence of the Ca$_3$Ru$_2$O$_7$ RIXS spectrum at \mbox{$T=25$~K} and \mbox{$E_i=2838$~eV}, covering the $t_{2g}$ multiplets and magnetic signal. The datasets were collected by varying the angle $\theta$ between the incoming photon polarization and the crystallographic $c$-axis (see Fig~\ref{fig:rixs}b) to cover almost grazing incidence at $\theta=10^\circ$ to $\theta=45^\circ$. This range corresponds to changing the in-plane momentum-transfer from \mbox{$Q=(1,0)$} to \mbox{$Q=(0,0)$}. The spectra are normalized to a flat region around 2.0~eV in order to simply account for the effects of self-absorption. The data at \mbox{$Q=(1,0)$} is overlaid in grey with spectra collected at other incidence angles for comparison. Here it can be seen that while the magnon clearly disperses from $\Gamma$, the higher energy signal associated with the $t_{2g}$ excitations shows no measurable shift in spectral weight.

The lack of any clear dispersion implies that the origin of the excitations in the $t_{2g}$ multiplet are local in nature and indeed arise from intra-ionic interactions. At the same time, the $t_{2g}$ multiplet does undergo a small decrease in intensity as the incident angle is increased. This is in marked contrast to the polarization dependency of Ca$_2$RuO$_4$, which shows a very strong intensity variation of the \mbox{$\mathrm{J}=2$} excitation, effectively doubling between $\theta=45^\circ$ and $\theta=10^\circ$~\cite{Gret19}. This is a hallmark of the strong tetragonal distortion in Ca$_2$RuO$_4$, which drives the condensation of \mbox{$\mathrm{J}=1$} excitations. In Fig.~\ref{fig:tdep}d) the integrated intensity of the Ca$_3$Ru$_2$O$_7$ excitations between 0.2 and 0.6~eV is plotted as a function of incidence angle and compared with the Ca$_2$RuO$_4$ \mbox{$\mathrm{J}=2$} intensity. The intensity of the Ca$_3$Ru$_2$O$_7$ signal is clearly devoid of such enhancement. Given that the magnitude of $\xi$ is unlikely to change by any significant amount between the two systems, it is clear that the crystal field distortion is smaller in Ca$_3$RuO$_7$, and that the intra-ionic spin-orbit transitions strongly interact with the overwhelming electron-hole continuum.

\section{Discussion}

Our experimental findings reveal two important features of Ca$_3$Ru$_2$O$_7$. First, clearly dispersing ferromagnetic excitations were identified below $T_{\mathrm{S}}$, in excellent agreement with inelastic neutron scattering results. Second, unlike the sharp multiplets of Ca$_2$RuO$_4$, the RIXS spectra of Ca$_3$Ru$_2$O$_7$ shows a broad asymmetric feature with two energy scales---a metallic continuum response, and excitations arising from damped ionic correlations in the $t_{2g}$ band.

The agreement between INS and RIXS validates our magnon observation and emphasizes the complementary nature of the two techniques. Although the energy resolution of INS is superior to RIXS it lacks the count rates at higher-energy loss (see zone-boundary data in Fig.~\ref{fig:mag}a). With this complete dataset at hand we can with confidence report a magnon gap of $\sim$8~meV and a zone-boundary energy of $\sim$37~meV in Ca$_3$Ru$_2$O$_7$. We observe that the nearest-neighbor exchange in Ca$_3$Ru$_2$O$_7$ is ferromagnetic \mbox{($J=-3.75$~meV)} instead of AFM in Ca$_2$RuO$_4$ (5.8~meV), a hallmark of the correlated metallic state. In Fig.~\ref{fig:mag}b) we have included the peak position of the Ca$_2$RuO$_4$ magnon excitation at $\Gamma$ as determined from Fig.~\ref{fig:comp}, as well as the model spin wave from Ref.~\onlinecite{Jain17} as a dashed white line, which clearly show the different magnetic dynamics of the two systems. More notable is the drastic reduction of the tetragonal SIA term \mbox{$E=5.5$~meV} when compared with Ca$_2$RuO$_4$ (22.75~meV). The primary driver behind the reduction in $E$ is likely the weaker $\Delta$ tetragonal crystal field in Ca$_3$Ru$_2$O$_7$ due to reduced RuO$_6$ distortions~\cite{Brad98, Yosh05}. It is therefore of interest that the magnon gap appears only below $T_{\mathrm{S}}$ and \emph{not} at $T_{\mathrm{N}}$, demonstrating directly the strong coupling between the structure and the magnetic moments. More specifically, these results show that the transition between the AFM-$a$ and AFM-$b$ states is associated with a marked change in spin anisotropy.
 
Our magnon results also give important insight into the field of RIXS studies of Ru-based compounds. In a recent O $K$-edge RIXS experiment on Ca$_3$Ru$_2$O$_7$ the magnon mode we identify at the Ru $L_3$-edge is not observed, despite the better energy resolution~\cite{Arx20}. Although it is well known that single magnons are generally silent at the O $K$-edge due to the lack of SOC in the $1s$ core-hole~\cite{Biso12}, special cases exist where this does not hold. A recent example includes the structurally related $5d$-electron iridate systems Sr$_2$IrO$_4$ and Sr$_3$Ir$_2$O$_7$, where both magnons and bi-magnons could be probed thanks to the presence of strong SOC within the Ir $5d$ $t_{2g}$ orbitals~\cite{Lu18}. This suggests that even though SOC is present in the Ru $4d$ $t_{2g}$ orbitals it may not be strong enough to facilitate single spin-flip excitations, leaving only bi-magnons as an option at the O $K$-edge. In view of this observation, it is interesting to note that O $K$-edge RIXS reports a feature at 55~meV in Ca$_3$Ru$_2$O$_7$~\cite{Arx20}, a mode that we are not able to clearly identify at the $L_3$-edge, presumably due to the dominant magnon intensity.

We now move our discussion to the electronic excitations associated with the $t_{2g}$ multiplet. It poses a challenge to extract the underlying structure in the RIXS spectrum of any metallic system due to the lack of sharp defining features. Here however, via analysis of the incidence-energy and temperature dependencies studies of the RIXS response of the system, we demonstrate that intra-atomic excitations survive in the metallic state. The comparison with Ca$_2$RuO$_4$ in Fig.~\ref{fig:comp} is then striking, as it shows that the RIXS spectra of Ca$_3$Ru$_2$O$_7$, while heavily damped by the overlapping electron-hole continuum, nevertheless includes fingerprints of spin-orbit excitations and Hund's multiplets that are hallmarks of the ionic model. The ionic SOC $\xi$ and Hund's rule $J_{\mathrm{H}}$ terms that define these excitations are unlikely to be meaningfully different between the two systems. As such, these results indicate that the Ca$_3$Ru$_2$O$_7$ $t_{2g}$ multiplet structure is reconfigured primarily by a smaller $\Delta$ term (which we note also drives the reduction of the tetragonal SIA spin-wave parameter). This observation is also supported by the optical conductivity, which remains featureless even at 300~meV~\cite{Lee07}, excluding a simple band structure origin. At the same time, it is unclear why the spin-orbit excitation would gain intensity in the paramagnetic state as Fig.~\ref{fig:tdep}a) shows. One possibility is that this is a consequence of a change in the electronic structure, although the energy scale here is much larger than reported in Raman and optical conductivity experiments~\cite{Liu99, Lee07}. Further investigation is needed to resolve this, including realistic dynamical mean-field theory (DMFT) calculations where local exchanges and band structure are treated equally.

\section{Conclusion}

We have reported upon extensive Ru $L_3$-edge RIXS studies of the correlated multiband bilayer system Ca$_3$Ru$_2$O$_7$. A well defined magnetic excitation is observed in excellent agreement with inelastic neutron scattering results, further forging an important link between these two experimental techniques. At higher energies we discover a broad asymmetric $t_{2g}$ excitation that is in stark contrast to the sharp multiplets in Ca$_2$RuO$_4$. However, the behavior of this feature in response to temperature, incidence-energy, and momentum-transfer suggest that the vestiges of ionic multiplets remain present in the metallic state of Ca$_3$Ru$_2$O$_7$, albeit heavily mixed with the electron-hole continuum. Our results show that RIXS at the $4d$-electron $L$-edges is highly sensitive to the presence of on-site atomic interactions, providing an essential view into the interactions that underlie correlated behavior in multiband metallic systems.

\acknowledgments{We kindly thank A.~Boris, K.~Rabinovic, G.~Khaliullin and H.~Liu for fruitful discussions. The project was supported by the European Research Council under Advanced Grant No. 669550 (Com4Com). We acknowledge DESY (Hamburg, Germany), a member of the Helmholtz Association HGF, for the provision of experimental facilities.}

\end{document}